\documentclass[a4paper]{jpconf}

\usepackage{graphicx,color}
\usepackage{amssymb}
\usepackage{amsmath}

\begin{document}

\renewcommand{\Im}[0]{\mathrm{Im}\,}
\renewcommand{\Re}[0]{\mathrm{Re}\,}
\renewcommand{\tilde}[0]{\widetilde}

\newcommand{\comment}[1]{!!! \textbf{#1} !!!}
\newcommand{\ie}[0]{i.e.\@\xspace} 
\newcommand{\eg}[0]{e.g.\@\xspace}

\newcommand{\om}[0]{\omega}
\newcommand{\gammab}[0]{\bar{\gamma}}
\newcommand{\Ep}{E_\mathrm{p}}
\newcommand{\omb}[0]{\bar{\omega}}
\newcommand{\Ap}[0]{A_\mathrm{p}}
\newcommand{\Ae}[0]{A_\mathrm{e}}
\newcommand{\me}[0]{\mathrm{e}}
\newcommand{\mi}[0]{\mathrm{i}}
\newcommand{\md}[0]{\mathrm{d}}
\newcommand{\nag}{\phantom{\dag}}
\newcommand{\op}{\hat{p}}
\newcommand{\ox}{\hat{x}}
\newcommand{\on}{\hat{n}}

\newcommand{\las}[0]{\langle}
\newcommand{\ras}[0]{\rangle}
\newcommand{\la}[0]{\left\las}
\newcommand{\ra}[0]{\right\ras}
\newcommand{\ket}[1]{\left|#1\ra}
\newcommand{\bra}[1]{\la#1\right|}
\newcommand{\sket}[1]{|#1\ras}  
\newcommand{\sbra}[1]{\las#1|} 
\newcommand{\braket}[2]{\la#1\left.\right|#2\ra}
\newcommand{\sbraket}[2]{\las#1\left.\right|#2\ras}

\newcommand{\dbraket}[2]{\la\la#1\left.\right|#2\ra\ra}

\title{Transport through a vibrating quantum dot: Polaronic effects}

\author{T Koch\dag, J Loos\ddag, A Alvermann\dag, A R Bishop\S,  
and H Fehske\dag}

\address{\dag\ %
  Institute of Physics, Ernst-Moritz-Arndt University Greifswald, 
17487 Greifswald, Germany
}
\address{\ddag\ %
  Institute of Physics, Academy of Sciences of the Czech Republic, 16200
Prague, Czech Republic
}

\address{\S\ %
  Theory, Simulation and Computation Directorate, 
Los Alamos National Laboratory, Los Alamos, New Mexico 87545, USA
}

\ead{holger.fehske@physik.uni-greifswald.de}

\date{\today}
\begin{abstract}
We present a Green's function based treatment
of the effects of electron-phonon coupling on transport
through a molecular quantum dot in the quantum limit.
Thereby we combine an incomplete variational Lang-Firsov approach
with a perturbative calculation of the electron-phonon self energy
in the framework of generalised Matsubara Green functions and
a Landauer-type transport description. Calculating the
ground-state energy, the dot single-particle spectral function
and the linear conductance at finite carrier density,
we study the low-temperature transport properties of the
vibrating quantum dot sandwiched between metallic leads in the
whole electron-phonon coupling strength regime. We discuss corrections
to the concept of an anti-adiabatic dot polaron and show
how a deformable quantum dot can act as a molecular switch.
\end{abstract}

\section{Introduction}

Recent progress in nanotechnology allows for the fabrication of electronic devices with organic molecules as the active elements, which may constitute an alternative to conventional semiconductor technology in the search for further miniaturisation. 
The basic example for such a device is a single organic molecule contacted with metallic leads.
The molecule can be described as a quantum dot, i.e. as a system of finite size that is coupled to macroscopic charge reservoirs.
The electronic transport properties of such molecular electronic components depend on the geometry of the molecule and the properties of the molecule-lead contact, as well as on the average charge of the molecule~\cite{GRN07}. 
Since the molecule is so small, quantisation of energy levels becomes important.
Furthermore, the molecule is susceptible to specific structural changes in the presence of charge carriers. The energy of these deformations or vibrations can be comparable to the kinetic energy of passing particles, whose mobility may therefore be substantially affected.
This process seems to play a fundamental role in the observed non-linear behaviour of basic molecular devices, including negative differential resistance and hysteresis \cite{LD07}. 
With strong coupling to vibrational degrees of freedom, current rectification and fast current switching may be realised~\cite{Bra06}.

\section{Model}

 To investigate transport in a molecular quantum dot, we consider the model Hamiltonian
\begin{equation}\label{hamiltonian}
 H  =  \sum_{k,a} E_k^{}c_{ka}^{\dag} c_{ka}^{\nag} + \Delta d^\dag d^{\nag}  -g \om_0 ( b^{\dag} + b) d^\dag d  + \om_0  b^{\dag} b -\frac{t_d}{\sqrt{N}} \sum_{k,a} \left(d^{\dag}c_{ka}^{\nag}+c_{ka}^{\dag}d\right) \,.
\end{equation}  
Here $c_{ka}^{\dag}$ ($c_{ka}^{\nag}$) are creation (destruction) operators of non-interacting electrons with energy $E_k$ ($k=1, \dots, N$) in the left and right lead ($a=l,r)$.
 We assume that the leads are semi-infinite one-dimensional chains with a semi-elliptical density of states $\varrho(\xi)= \frac{1}{N}\sum_k \delta(\xi-E_k) = (2/\pi W^2) \sqrt{W^2-\xi^2} \, \Theta(W^2-\xi^2)$, where $W$ is the half bandwidth. 
The quantum dot is modelled by a single level $\Delta$ with fermionic operators $d^{(\dag)}$. 
Working with spinless fermions, we take a large local Coulomb 
repulsion at the quantum dot for granted. 
To describe the deformation of the molecule, an electron at the dot couples via a Holstein-like term ($\propto g$) to a local phonon mode $b^{(\dag)}$ of energy $\omega_0$.
The last term in (\ref{hamiltonian}) allows for the dot-lead particle transfer ($\propto t_d$).

In analogy to Holstein's small polaron theory \cite{Ho59a}, we expect -- for sufficiently large electron-phonon (EP) coupling $g$ and frequency $\omega_0$ -- the formation of a polaron-like state at the dot, corresponding to an electron 
with an accompanying phonon cloud. To account for this effect, a generalised Lang-Firsov-transformation $\tilde H =UHU^{\dag}$ with $U=\exp\{\tilde g (b^{\dag}-b)d^{\dag}d \}$ and $\tilde g=\gamma g$ is applied, where $\gamma\in[0,1]$ is a variational parameter. The transformed Hamiltonian reads
\begin{eqnarray} \label{hamiltonian2}
  \tilde{H} & = &\sum_{k,a} E_k^{} c_{ka}^{\dag} c_{ka}^{\nag}+\tilde \Delta d^\dag d^{\nag} -C_{d}d^{\dag} d + \om_0     b^{\dag} b  - \sum_{k,a}\left(C^{\phantom{\nag}}_{t}d^{\dag}c_{ka}^{\nag}+C^{\dag}_{t}c_{ka}^{\dag}d\right) \;,
\end{eqnarray}
with the renormalised dot-level $\tilde \Delta = \Delta -\varepsilon_p\gamma(2-\gamma)$ and the polaron binding energy $\varepsilon_p=g^2\omega_0$. 
As $\gamma$ grows from zero to one, our approach interpolates between a weak coupling ansatz ($\gamma=0$, $\tilde H=H$), and the complete Lang-Firsov-transformation ($\gamma=1$) restricted to large phonon frequencies and strong EP coupling,
when the direct EP coupling via $C_d=g\omega_0(1-\gamma)(b^\dag+b)$ is replaced by a phonon-affected dot-lead transfer term $C_t=(t_d/\sqrt{N})\exp\{-\tilde g(b^{\dag}-b)\}$.
In this way, we are able to describe the system for a large range of parameter values.
Note that the effects of a finite Coulomb interaction might be 
included by considering instead of the Hamiltonian~(\ref{hamiltonian}) 
Hubbard/Anderson-Holstein-type models~\cite{FWLB08,MR09}. Here the influence of the phonons is mainly to suppress the repulsion between the electrons at the molecular orbitals~\cite{FILTB94}. When the energy scales set by the Coulomb and EP interactions become comparable bipolaronic states may form at the quantum dot~\cite{FWLB08}. 
\section{Theoretical approach}
Our main interest lies in the single particle spectrum at the dot and the linear conductance in the case of vanishing voltage bias between the leads. Both of these quantities can be obtained from the polaronic spectral function $A_{dd}(\om)=\lim_{\delta\to 0^{+}}[\Im G_{dd}^R(\om+\mi\delta)]$, where the retarded Green function $G^{R}_{dd}$ corresponds to the (polaronic) operators $d^{(\dag)}$ in the transformed Hamiltonian in equation~\eqref{hamiltonian2}.
We base our calculations on the equations of motion of generalised Matsubara Green functions in 
equilibrium~\cite{KB62,LKABF09}
\begin{equation} \label{gdd}
  G_{dd}(\tau_1,\tau_2 ;\{V\})=-\frac{1}{\langle S \rangle} \langle \mathcal{T}_{\tau}d(\tau_1)d^{\dag}(\tau_2 )S  \rangle\,.
\end{equation}
The mean value and the time dependences in (\ref{gdd}) are determined by $\tilde H-\mu \hat{N}$, where $\mu$ is the equilibrium chemical potential of the system and $\hat{N}$ denotes the particle
number operator. The S-matrix 
\begin{equation}
\label{smatrix}
S= \mathcal{T}_{\tau}\exp\left\{-\int_{0}^{\beta}\md \tau\, V^{\nag}_{t}(\tau)C^{\nag}_{t}(\tau)+\bar{V}^{\nag}_{t}(\tau)C^{\dag}_{t}(\tau)+V^{\nag}_{d}(\tau)C^{\nag}_{d}(\tau)\right\}
\end{equation}
describes the coupling to the components of a fictitious external potential $\{V\}$ .
\subsection{Dot spectral function}
Starting from the equations of motion for the dot Green function $G_{dd}$ and 
the lead-dot transfer Green function $G_{cd;ka}$, we express the occuring 
interaction terms by functional derivatives with respect to 
$\{V\}$ \cite{KB62}, e.g., 
\begin{eqnarray}\label{intterm}
   \frac{\langle \mathcal{T}_{\tau}  C_{t}^{\nag} (\tau_1)  c_{ka}^{\nag}(\tau_1) d^{\dag}(\tau_2)S[V] \rangle }{\langle S[V] \rangle }   = 
	-  \bar C_{t} (\tau_1;\{V\}) \, G_{cd;ka}(\tau_1,\tau_2;\{V\})  + \, \frac{\delta G_{cd;ka}(\tau_1,\tau_2;\{V\})}{\delta V_{t}(\tau_1)} \,,
\end{eqnarray}
with the interaction coefficients $\bar C_{t}^{(\dag)}(\tau;\{V\})\equiv \langle\mathcal{T}_{\tau} C_{t}^{(\dag)}(\tau)S \rangle/\langle S \rangle$.
The resulting coupled equations lead to a functional differential equation for the polaronic self energy $ \Sigma_{dd}^{\nag} ( \tau_1,\tau_2;\{V\}) =  G_{dd}^{(0)-1} ( \tau_1,\tau_2)-  G_{dd}^{-1} ( \tau_1,\tau_2;\{V\})$ (see equation~(17) 
in~\cite{LKABF09}), which can be evaluated in a two-step process. First, we neglect terms with functional derivatives of $ \Sigma_{dd}^{\nag}$. We then insert the result, $\Sigma_{dd}^{(1)}$, into these derivatives and only keep terms up to second order in $\bar C_{t}^{(\dag)}$. In this way, we find 
\begin{eqnarray} \label{sigma}
& &\hspace*{-1cm}\Sigma_{dd}^{(2)} ( \tau_1,\tau_2;\{V\})\; =  \;
  - \bar{ C}_d (\tau_1;\{V\}) \delta[\tau_1-\tau_2] + \sum_{k,a} \bar{ C}_{t}^{\nag} (\tau_1;\{V\}) G_{cc;\,ka}^{(0)}(\tau_1,\tau_2) \bar{ C}_{t}^\dag (\tau_2;\{V\}) \\
& & \hspace*{2cm}+\sum_{k,a}G_{cc;\,ka}^{(0)}(\tau_1,\tau_2) 
  \left[ \frac{1}{\langle S\rangle}\langle \mathcal{T}_\tau C_{t}^{\nag}(\tau_1) C_{t}^{\dag}(\tau_2) S\rangle - \bar{C}_{t}^{\nag}(\tau_1;\{V\})\bar{C}_{t}^{\dag} (\tau_2;\{V\})\right] \nonumber\\
  & &\hspace*{2cm} + \, G_{dd}^{(1)}(\tau_1,\tau_2) \left [ \langle \mathcal{T}_\tau  C_d(\tau_1)C_d(\tau_2) \rangle - \bar{ C}_d (\tau_1;\{V\})\bar{ C}_d (\tau_2;\{V\}) \right] \nonumber \;.
\end{eqnarray}
Here  $G_{dd}^{(1)}$ denotes the Green function determined by the first order self energy and $G_{cc;\,ka}^{(0)}$ is the free Green function of the lead states. We then let $\{V\}\to0$ and calculate the correlation functions of the interaction coefficients assuming an independent Einstein oscillator. After Fourier transformation and summation over bosonic Matsubara frequencies, the retarded self energy follows in the low-temperature approximation 
$\beta\om_0\gg 1$ as~\cite{LKABF09}  
\begin{eqnarray}\label{sigmatwo}
& & \hspace{-0.5cm}\Sigma_{dd}^{(2)}(\om+\mi\delta) \quad = \quad 2\,t_{d}^2\, e^{-\tilde g^2}\int_{-W}^{W} \md \xi\, \varrho(\xi)\,\frac{1}{\om+\mi\delta-(\xi- \mu)} \\
& & \hspace{0.3cm} + \,2\,t_{d}^2\,e^{-\tilde g^2}\sum_{s\geq 1}\frac{(\tilde g^2)^{s}}{s!}  \int_{-W}^{W} \md \xi\, \varrho(\xi) \left(\frac{n_{F}(\xi- \mu)}{\om+\mi\delta-(\xi- \mu)+s\om_0} + \frac{1-n_{F}(\xi- \mu)}{\om+\mi\delta-(\xi- \mu)-s\om_0}\right) \nonumber\\
& & \hspace{0.3cm} +\left [(1-\gamma)g\om_0\right ]^2\int_{-\infty}^{+\infty}\md \om' A^{(1)}_{dd}(\om')\left(\frac{n_{F}(\om')}{\om+\mi\delta-\om'+\om_0}+ \frac{1-n_{F}(\om')}{\om+\mi\delta-\om'-\om_0}\right)\nonumber \;.
\end{eqnarray}
Within our iterative scheme, the corresponding spectral function $A_{dd}^{(2)}\equiv A_{dd}$ is evaluated in a two step process using
\begin{equation}\label{specfunc}
  A_{dd}^{(n)}(\om) 
    =  - \frac{1}{\pi} \lim_{\delta\to0^+}  \frac{\Im\Sigma_{dd}^{(n)}(\om+\mi\delta)}{\left [\om + \mu -\tilde \Delta-\Re\Sigma_{dd}^{(n)}(\om+\mi\delta)\right ]^2+\left [\Im\Sigma_{dd}^{(n)}(\om+\mi \delta)\right ]^2}\;,
\end{equation}
where the first order self energy $\Sigma_{dd}^{(1)}$ is given by the first two terms in (\ref{sigmatwo}). The self energy $\Sigma_{dd}^{(2)}\equiv \Sigma_{dd}^{}$ accounts for multi-phonon processes as well as finite particle densities. As $g\to0$, our model~(\ref{hamiltonian}) reduces to the Fano-Anderson Hamiltonian for a rigid impurity in a 1D lattice. Then only the first term in (\ref{sigmatwo}) remains and gives the exact self energy. For finite EP coupling, the spectrum contains multiple phononic side bands. If the condition $\om_0<W-|\mu|$ is fulfilled, these bands overlap and $\Im \Sigma_{dd}\neq 0$ along the whole $\om$-axis. Otherwise, a localised polaron-like state may exist in the intervals where $\Im\Sigma_{dd}=0$.

For the numerical evaluation of the spectral function we keep $\delta$ in (\ref{specfunc}) as a small positive parameter ($\delta\lesssim 5\cdot 10^{-3}$).
This avoids the problematic evaluation of principal value integrals in $\Re\Sigma_{dd}$ for $\delta\to 0$.
The resulting spectral function fulfils the sum rule $\int A_{dd}(\om) \md\om=1$, which is preserved in our approximations.

\subsection{Ground state energy}
The variational parameter $\gamma$ is determined by minimisation of the ground state energy $E=\langle \tilde H -\mu \hat{N} \rangle$ with respect to $\gamma$. 
We identify the statistical averages in $E$ with expressions like equation~(\ref{intterm}), letting $\tau_1^{\phantom{}}\to \tau_2^{-}$ and $\{V\}\to 0$. As in the evaluation of $\Sigma_{dd}^{(1)}$, we neglect the functional derivatives in 
equation~(\ref{intterm}), so that $\langle C^{\phantom{\nag}}_{t}d^{\dag}c_{ka}^{\nag} \rangle \approx -\langle C^{\phantom{\nag}}_{t} \rangle G_{cd;ka}(\tau_1,\tau_2^{})\big |_{\tau_1^{\phantom{}}\to \tau_2^{-}} $.  
Applying the same approximation to the equations of motion of the Green functions, we determine their Fourier transforms in the complex plane to first order as 
\begin{eqnarray}
G_{cc;ka}(z)&\approx& G_{cc;ka}^{(0)}(z)+ \langle C^{\dag}_{t} \rangle \langle C^{\nag}_{t} \rangle \big [ G_{cc;ka}^{(0)}(z) \big ]^{2} G_{dd}(z)\;,\\
 G_{cd;ka}(z) &\approx& - \langle C^{\dag}_{t} \rangle G_{cc;ka}^{(0)}(z)G_{dd}(z)\;.
\end{eqnarray}
For the variation of $\gamma$ we can omit the constant lead energy related
to $G_{cc;ka}^{(0)}$. In equilibrium, $\langle b^{\dag}b\rangle$ and $\langle C_{d}\rangle$ vanish as $T\to0$, and we obtain
\begin{eqnarray}\label{htilde}
 \hspace*{-0.5cm}E & = &  -\,2\, t_d^2\,e^{-\tilde g^2} \int_{-W}^{\mu} \md  \xi \,\varrho(\xi)  \int_{0}^{+\infty} \md \om'  \, A_{dd}(\om') \left [ \frac{2}{\om' -(\xi-\mu)} + \frac{\xi-\mu}{(\om' -(\xi-\mu))^2} \right ] \nonumber\\
&& + \,2\, t_d^2\,e^{-\tilde g^2} \int_{\mu}^{W} \md  \xi \,\varrho(\xi)  \int_{-\infty}^{0} \md \om'  \, A_{dd}(\om') \left [ \frac{2}{\om' -(\xi-\mu)} + \frac{\xi-\mu}{(\om' -(\xi-\mu))^2} \right ] \nonumber\\
&&+\,(\tilde \Delta - \mu ) \int_{-\infty}^{0} \md  \om' A_{dd}(\om') \,.
\end{eqnarray}
Via $A_{dd}$, $\tilde \Delta$ and $\tilde g$, the energy $E$, given by equation~(\ref{htilde}), depends on $\gamma$ and can be used to determine the extremal variational parameter $\gamma_{min}$.

\subsection{Conductance}
We finally calculate the linear conductance $L$ starting from the Meir-Wingreen formula~\cite{MW92a}, which expresses the current $J$ through the dot in terms of the retarded non-equilibrium Green function $ G^R_{dd}$ of the electronic operators $d$ in (\ref{hamiltonian}).
For equal coupling to the left and right lead, and finite voltage bias $\Phi=-(\mu_l-\mu_r)/\me$, the current takes the simple form
\begin{eqnarray}\label{current}
J  = - \e\, t_d^2\,\int_{-W}^{W} \md \xi\, \varrho(\xi)\,[n_F(\xi-\mu_l) - n_F(\xi-\mu_r) ]\, \Im G_{ d d}^{R}(\xi)  \;.
\end{eqnarray}
Here $n_F(\xi)=(e^{\beta \xi}+1)^{-1}$ is the Fermi distribution function of the isolated leads in thermal equilibrium, at respective chemical potential $\mu_{l/r}$. 
For small voltage bias, i.e. $\mu_{l,r}=\mu\pm\delta \mu/2$, we can express the current as $J= - L \delta \mu/\me$. Then, the linear conductance $L=\lim_{\delta\mu\to0}\{-\me J/\delta\mu\}$ results from (\ref{current}) as 
\begin{equation}\label{coeffnf}
L=\me^2\pi\,t_d^2\,\int_{-W}^{W} \md \xi\, \varrho(\xi) \,[-n_F'(\xi-\mu)]\,A^e_{dd}(\xi-\mu)\;,
\end{equation} 
where the electronic spectral function $ A^e_{d d}(\om)=\lim_{\delta\to 0^{+}}[\Im G_{ d d}^R(\om+\mi\delta)]$ is now calculated in equilibrium. Based on the factorisation of the statistical averages of phonon and polaron variables, a relation between $ A^e_{dd}$ and the polaronic $A_{d d}$ has been derived in equation ($40$) of \cite{LHF06}:
\begin{eqnarray}\label{electronic}
 A^e_{dd}(\om) = e^{-\tilde g^2}  \sum_{s\geq 0}\frac{(\tilde g^2)^{s}}{s!} 
	\left [ A_{dd}(\om-s\om_0)\Theta(\om-s\om_0) + A_{dd}(\om+s\om_0)\Theta(-\om-s\om_0)\right ] \;.
\end{eqnarray}
Then, from (\ref{coeffnf}) and (\ref{electronic}), the linear conductance in the low-temperature approximation is
\begin{eqnarray}\label{coeff}
L= \me^2\pi\,t_d^2\, e^{-\tilde g^2} \varrho(\mu) A_{dd}(0)\;.
\end{eqnarray} 
As usual, it depends on the accessibility of dot states at the Fermi level.
In addition the conductance exhibits a Lang-Firsov renormalisation $\tilde{t}_d = t_d e^{-\tilde{g}^2/2}$ of the dot-lead transfer integral, in accordance with the results for the Holstein model.

\section{Numerical results and discussion}
With our choice of a semi-elliptical lead density of states, 
the non-interacting system ($\epsilon_p=0$) is translational invariant if the dot-lead transfer $t_d$ is equal to the value $t=W/2$ for electron hopping along the leads.
In the following, we fix $t=1$ and set $t_d=0.25$, representing the 
weak dot-lead coupling case.
\begin{figure}[t]
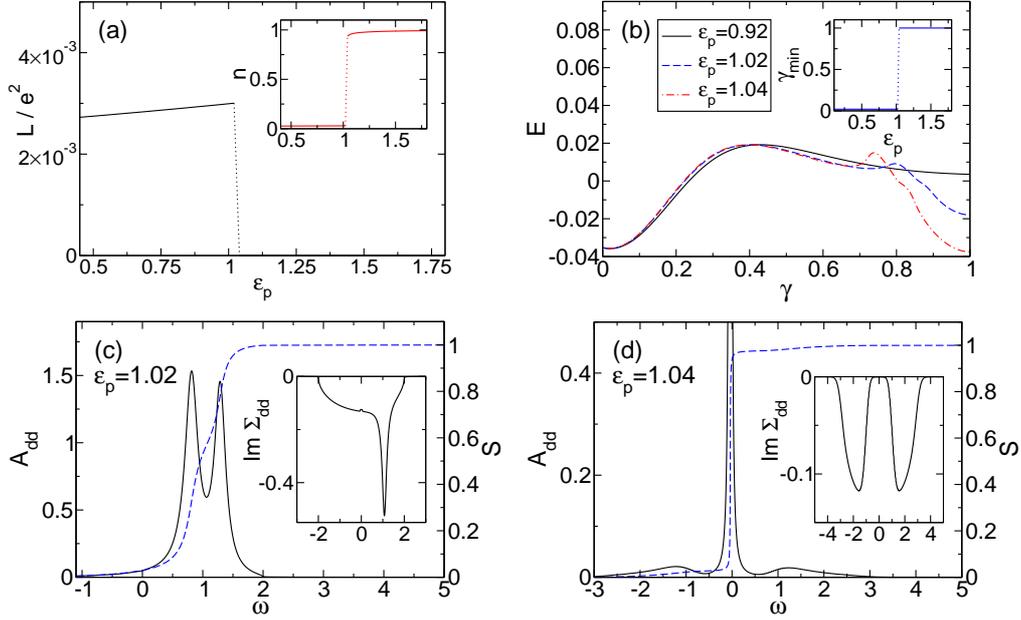

\begin{center}
\hspace{-0.6cm}\includegraphics[width=0.365\textwidth]{fig1a}\hspace{0.85cm}
\includegraphics[width=0.369\textwidth]{fig1b}\\[0.2cm]
\includegraphics[width=0.405\textwidth]{fig1c}\hspace{0.2cm}
\includegraphics[width=0.405\textwidth]{fig1d}
\end{center}
\caption{For model parameters $\om_0=0.05$, $\Delta=1$, $t_d=0.25$, $\mu=0$. Panel (a): Linear conductance $L$ and occupation of the dot $n$ (inset) as functions of $\varepsilon_p$. At $\varepsilon_p^{c}=1.03$ an abrupt transition takes place. Panel (b): Dot energy $E$ as a function of $\gamma$ for several EP coupling strengths. At $\varepsilon_p^{c}$, a second minimum at $\gamma=1$ becomes the global minimum, causing a jump in the extremal parameter $\gamma_{min}$ (inset). Panels (c) and (d): Spectral function $A_{dd}(\omega)$, integrated spectral weight $S(\om)=\int_{-\infty}^{\om} \md \om'\, A_{dd}(\om')$ and imaginary part of the self energy  $\Sigma_{dd}(\omega)$  for $\varepsilon_p$ in the vicinity of the transition, showing the sudden formation of a long-living polaron-like state.}
\label{fig1}
\end{figure}
\subsection{Adiabatic regime}
Let us first consider the adiabatic regime ($\om_0=0.05 < t_d$), where the phononic time-scale is much slower than the electronic time-scale and the deformation of the dot adjusts quasi-statically to the average electronic occupation. We set $\mu=0$ and $\Delta=1$, so that without EP coupling ($\varepsilon_p=0$) the dot simply acts as an impurity leading to scattering. For these parameters the conductance $L=3\cdot10^{-3}$ is small, i.e. far below its value $L\approx0.16$ for $\Delta=0$ (see \fref{fig1}a). As $\varepsilon_p$ increases, $L$ and the dot occupation $n=\langle d^{\dag}d\rangle$ grow only marginally. At a critical EP coupling $\varepsilon_p^c=1.03$ a sudden transition to $n\approx 1$ takes place, resulting in a drop of $L$ by eight orders of magnitude. This behaviour can be traced back to a jump in $\gamma_{min}$, as \fref{fig1}b shows. For small $\varepsilon_p$, $E(\gamma)$ has a single minimum at $\gamma_{min}=0.02\ll1$. Hence there is almost no renormalisation of $\Delta$ or $t_d$ and $L$ is nearly independent of $\varepsilon_p$. However, the third term in (\ref{sigmatwo}) contributes and the single peak in $A_{dd}^{(1)}$ is added to $\Im\Sigma_{dd}$, resulting in two maxima of equal spectral weight around $\om=\Delta$ in $A_{dd}$ (see \fref{fig1}c). At $\varepsilon_p=\varepsilon_p^c$, when the gain in potential energy overcompensates the loss in kinetic energy, a second minimum of $E(\gamma)$ appears and becomes the global minimum. As a result $\gamma_{min}$ jumps to unity, which corresponds to a complete Lang-Firsov transformation. The dot level ($\tilde \Delta = \Delta -\epsilon_p=-0.03$) is shifted below the Fermi level  and the dot-lead transfer is reduced by a factor of $\exp\{-\tilde g^2\}$ with $\tilde g^2\geq 20$, so that $L$ vanishes. Because $\om_0\ll W$, the spectrum consists of overlapping phononic bands around a pronounced central peak, which signals the formation of a long-living polaron-like state at the quantum dot.
We know from variational approaches to the polaron problem, that the jump in $\gamma_{min}$ may be an artefact of our variational ansatz. In the Holstein model with EP interaction at every lattice site, the formation of a heavy polaron can occur as a sharp but always continuous crossover for small phonon frequencies. In contrast, a true phase transition from zero to finite dot occupation $n$ is found for a single electron at the quantum dot with $\Delta>0$ \cite{AF08b}. This phase transition becomes more pronounced as $\om_0$ gets smaller. In \cite{FWLB08}, similar behaviour is found in a generalised Holstein-Hubbard model with site-dependent potentials and EP couplings. Therefore, our results do not contradict the overall physical picture, and the variational Lang-Firsov transformation simulates the rapid adiabatic transition by a discontinuous change in $\gamma_{min}$.
\begin{figure}[t]
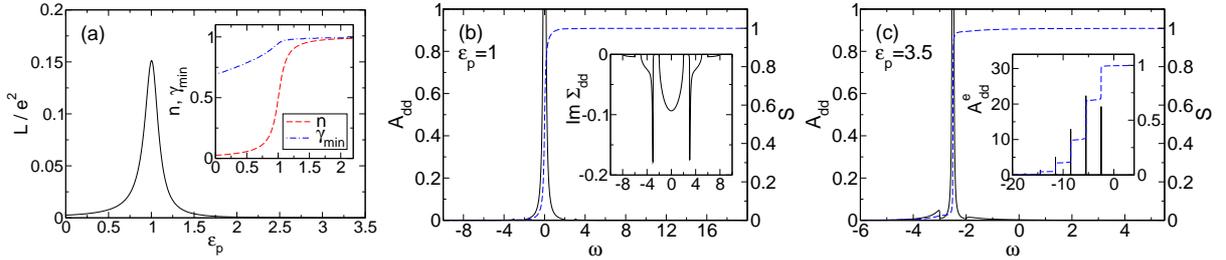

\begin{center}
\includegraphics[width=0.305\textwidth]{fig2a}
\includegraphics[width=0.335\textwidth]{fig2b}
\includegraphics[width=0.335\textwidth]{fig2c}
\end{center}
\caption{For model parameters $\om_0=3$, $\Delta=1$, $t_d=0.25$, $\mu=0$. Panel
(a): Because $E$ has a single minimum at $\gamma_{min}$ for all $\varepsilon_p$, a smooth crossover with $\gamma_{min}\geq 0.7$ occurs (inset). The linear conductance reaches a maximum at $\varepsilon_p\approx \Delta$, when the peak in $A_{dd}$ is shifted to the Fermi level (c.f. panel (b)). Panel (c): For large $\varepsilon_p$, a bound polaron-like state may exist in the intervals where $\Im\Sigma_{dd}=0$. Then the electronic spectrum exhibits a multi-peak structure.}
\label{fig2}
\end{figure}
\subsection{Anti-adiabatic regime}
In the anti-adiabatic regime with $\om_0=3\gg t_d$, the dot deformation adjusts instantaneously to the presence of an electron. As \fref{fig2}a shows, our approach is better suited to this regime: For $\mu=0$ and $\Delta=1$, $E(\gamma)$ has a single minimum $\gamma_{min}\geq 0.7$ for all EP coupling strengths and we find a smooth crossover to $n\approx 1$. At the quantum dot, both the renormalisation of $t_d$ and $\Delta$ affect the transport. Since the phonon number $g^2=\varepsilon_p/\om_0$ is small in the anti-adiabatic case, the shift of the dot level is the predominant effect. With increasing $\varepsilon_p$, the repulsive potential $\tilde \Delta$ is continuously lowered and the conductance $L$ grows, reaching almost the maximum of the $\Delta =0$ case when $\varepsilon_p\approx \Delta$. Here the peak in $A_{dd}$ is shifted to the Fermi level (see \fref{fig2}b) and we can speak of phonon-assisted transport. Because $\om_0>W$, the self energy features few non-overlapping phonon bands (see inset of \fref{fig2}b). For large $\varepsilon_p$, a bound polaron-like state forms when $\tilde \Delta$ is located in the intervals with $\Im\Sigma_{dd}=0$, as \fref{fig2}c shows for $\varepsilon_p=3.5$ and $\tilde\Delta\approx-2.5$. The electronic spectrum then exhibits the typical structure of multiple Poisson-weighted peaks (inset). Note that the delta peaks have finite width due to our numerical parameter $\delta>0$. 

The results show that, as for the Holstein model, the complete Lang-Firsov transformation is restricted to large phonon frequency and strong electron-phonon coupling. Away from this limit, $\gamma_{min}$ significantly differs from one, and our approach allows for important corrections in the regimes of weak coupling and moderate to small phonon frequencies. 
\begin{figure}[t]
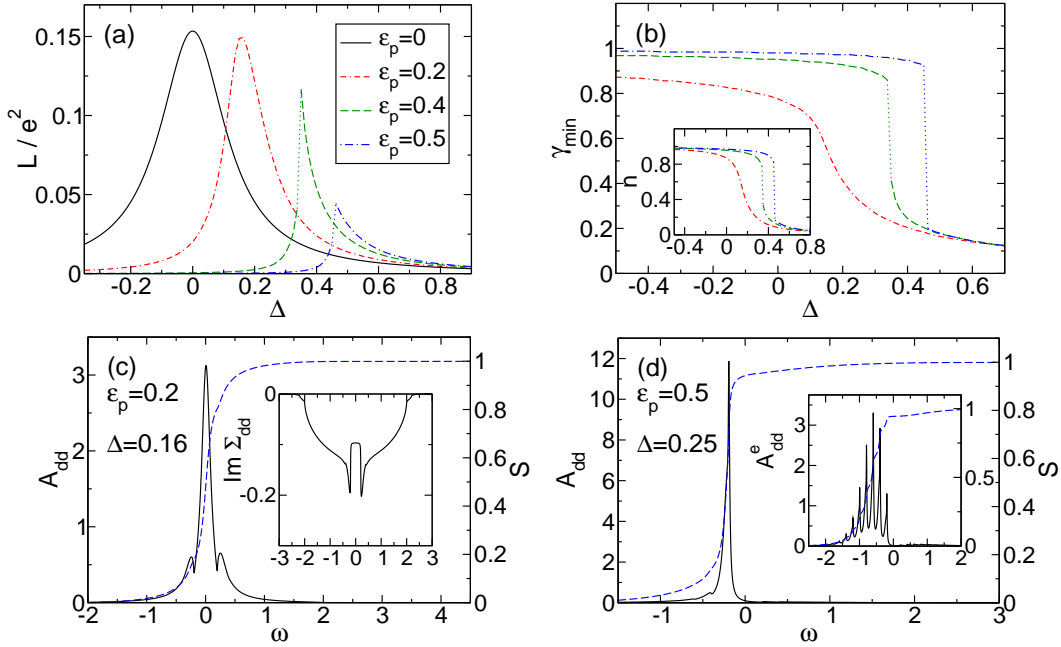

\begin{center}
\hspace{-1.1cm}\includegraphics[width=0.383\textwidth]{fig3a}\hspace{0.9cm}
\includegraphics[width=0.374\textwidth]{fig3b}\\[0.2cm]
\includegraphics[width=0.405\textwidth]{fig3c}\hspace{0.2cm}
\includegraphics[width=0.414\textwidth]{fig3d}
\end{center}
\caption{For model parameters $\om_0=0.2$, $t_d=0.25$, $\mu=0$. Panels (a) and (b): $L$ and $\gamma_{min}$ as a function of the quantum dot level. For small 
$\varepsilon_p=0.2$, we find a smooth crossover.  As $\varepsilon_p$ grows, a jump-like transition occurs and the dot acts as a molecular switch. Panel (c): Spectral function at the point of maximum conductance in the smooth crossover. Panel (d): Due to overlapping phononic bands, no bound state exists even for large $\varepsilon_p$.}
\label{fig3}
\end{figure}

\subsection{Intermediate phonon frequency}
Next we investigate the intermediate regime, where phononic and electronic energies become comparable ($\om_0=0.2$). We consider an experimentally relevant situation by keeping the EP coupling fixed and varying the dot level (gate voltage). In the case of a rigid quantum dot ($\varepsilon_p=0$), scattering off the dot potential reduces the conductance. Therefore, in \fref{fig3}a, $L$ exhibits a maximum at $\Delta=0$ and decreases symmetrically as $|\Delta|$ grows. For finite EP coupling $\varepsilon_p=0.2$, this maximum is shifted to $\Delta>0$, where the repulsive dot potential is compensated by the EP interaction. The spectral function then shows a peak at the Fermi level and two phonon satellites (see \fref{fig3}c).
As we see from \fref{fig3}b, the optimal variational parameter is a continuous function of $\Delta$ taking values from 0.1 to 0.9. Therefore, the effective renormalisation of $\tilde{t}_d$ and $\tilde \Delta$ depends, via $\gamma_{min}$, on the dot level $\Delta$ itself. In contrast to the result for a complete Lang-Firsov transformation with fixed $\gamma=1$ (cf. figure 5 of \cite{GNR06}), the conductance maximum is shifted by less than $\varepsilon_p$, and $L$ decreases asymmetrically away from this point. In accordance with Galperin {\it et al} \cite{GNR06} and Mitra {\it et al} \cite{MAM04}, we find no phonon side peak in $L(\Delta)$. 
If $\varepsilon_p$ is set to larger values ($\varepsilon_p=0.4$ and $0.5$), 
a transition in the dot occupation appears which is again related to 
a jump in $\gamma_{min}$. 
La Magna and Deretzis \cite{LD07} found a similar behaviour: for intermediate phonon frequencies and strong EP coupling their variational ansatz showed bistabilities causing a sudden occupation of the quantum dot (cf. figs. 2c and 2d in \cite{LD07}). 
As the dot level falls below a critical value, the charging of the quantum dot is accompanied by a drop in the conductance. In this way the quantum dot acts as a simple molecular switch. Now the spectrum shows a pronounced maximum at $\tilde \Delta$, which is below the Fermi level. Note that, due to overlapping phonon bands, no bound state exists even for large $\varepsilon_p$ (\fref{fig3}d). 
\begin{figure}[b]
\begin{center}
\includegraphics[width=0.8\textwidth]{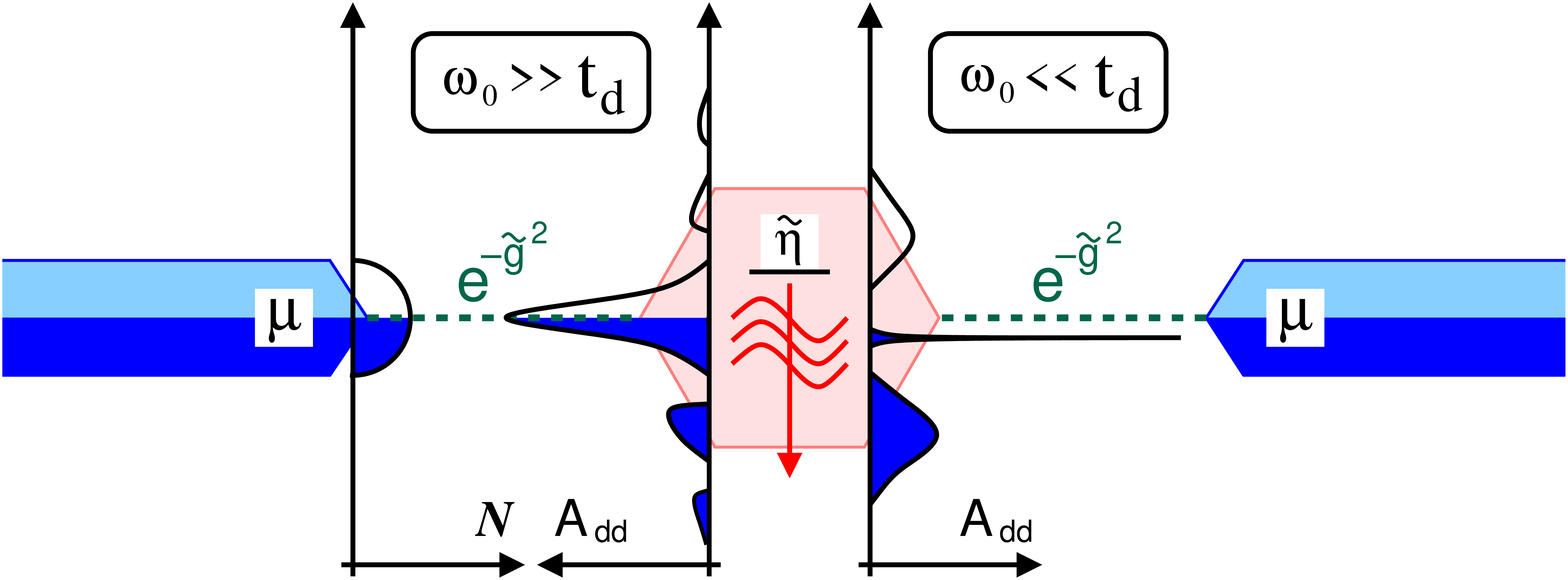}
\end{center}
\caption{Sketch of the basic polaronic effects on transport through a vibrating quantum dot. In addition to a renormalisation of the dot-lead transfer integral, in the anti-adiabatic regime, a continuous lowering of the dot potential towards the Fermi level enhances the transport. In the adiabatic regime, the sudden formation of a quasi-localised polaron state causes a drop of the conductance.}
\label{fig4}
\end{figure}
\section{Conclusion}

In the present contribution we studied transport through a deformable quantum dot, realised e.g. by an organic molecule in contact to leads.
Our treatment is based on Green's functions equations, with
the dot Green's function as the central quantity of our calculation.
EP interaction significantly affects the shape of the associated dot spectral function.
Since at low temperature the conductance of the quantum dot is
determined by the spectral weight close to the Fermi energy, it changes accordingly.
The basic effect, with dramatic consequences for the conductance, is the formation of a "localised" polaron-like dot state.
Similar to the Holstein polaron, this effect can be captured in a 
Lang-Firsov approach.
The virtue of our analytical approach lies in the variational determination of the Lang-Firsov parameter. 
This allows us to account for the basic polaronic effects both in the adiabatic and anti-adiabatic regimes, which are distinguished by their different influence on the spectral function and, accordingly, on the conducteance (briefly summarised in \fref{fig4}). 
Particularly for comparable phonon and electron time-scales we find interesting physical behaviour, which motivates further investigation of the quantum dot system, e.g. with respect to applications as a current switch.

The limitations of our study suggest two directions for improvements.
First, the inclusion of Coulomb interaction at the dot.
Many interesting effects we present here arise from the competition between electron-phonon interaction and a repulsive dot potential.
Coulomb interaction gives rise to an effective repulsive dot potential,
whose strength depends on the dot population. 
This may allow for even stronger non-linear behaviour, 
as then the transition between an effectively attractive or repulsive dot depends on the dot population through both interaction mechanisms.
Second, all our approximations neglect correlations between phononic and polaronic degrees of freedom (note however that the Lang-Firsov transformation introduces strong correlations between the phononic and electronic degrees of freedom).
In this way, the variational Lang-Firsov transformation can be used to map the entire problem approximately onto a purely electronic problem, as in reference~\cite{LD07}.
This reference reports results similar to ours,
e.g. for the dot population, but also for articifial bistabilities corresponding to different local minima of the groundstate energy, which are a common feature of these variational approaches.
As a first step towards the reintroduction of electron-phonon correlations
our treatment includes the weak-coupling contribution to the self-energy,
which partly accounts for retardation of electron-phonon interaction at finite phonon frequencies.
A detailed comparison of our results to those presented in reference~\cite{LD07} has to be given elsewhere, and the improvement of our treatment along the lines indicated is the subject of future work.

{\it Acknowledgements.} We are grateful to one referee for bringing 
reference~\cite{LD07} to our attention. H.F. acknowledges the hospitality
at the Institute of Physics, Czech Academy of Sciences, Prague. 
\section{References}
%

\end{document}